\documentclass[twocolumn]{aastex63}
\usepackage{amsmath,amstext}
\usepackage[T1]{fontenc}
\usepackage[figure,figure*]{hypcap}
\usepackage{multirow}
\usepackage[T1]{fontenc} 
\usepackage[utf8]{inputenc} 
\usepackage{float}
\newcommand{\ba}{\begin{eqnarray}}
\newcommand{\ea}{\end{eqnarray}}
\usepackage{hyperref}
\usepackage{lineno}
\usepackage{amssymb}

\graphicspath{{./}{figures/}}

\shorttitle{Parking lot planet}
\shortauthors{Best, Correa, Espinoza}
\received{April 1, 2022}

\begin{document}

\title{The Parking Lot Planet}

\correspondingauthor{Marcy Best}
\email{mbest@uc.cl}

\author[0000-0001-8361-9463]{Marcy Best}
\affiliation{Instituto de Astrofísica, Pontificia Universidad Católica de Chile, Santiago, Chile}
\affiliation{Institute for Planetary Architecture \& Design, Magrathea}

\author[0000-0002-4365-2258]{Fernanda Correa}
\affiliation{Instituto de Astrofísica, Pontificia Universidad Católica de Chile, Santiago, Chile}

\author[0000-0001-9480-8526]{Juan Ignacio Espinoza}
\affiliation{Instituto de Astrofísica, Pontificia Universidad Católica de Chile, Santiago, Chile}

\begin{abstract}

We give conditions for an exoplanetary system to function as an ideal amusement park/vacation resort (with its separate parking lot, of course); in case of massive human interplanetary colonization. Our considerations stem from the fact that an amusement park needs a parking lot of roughly the same surface area, thus the best option for its construction would be a system with at least 2 planets close to each other for easy tourist transportation.  We also discuss the likelihood of finding such a system out there to cut down on construction costs.



\end{abstract}


\section{Introduction} \label{introduction}

The newly reached goal of 5,000 confirmed exoplanets \footnote{\href{https://exoplanetarchive.ipac.caltech.edu/}{https://exoplanetarchive.ipac.caltech.edu/}} opens up the possibility for humanity to give each planet a distinct purpose, with a huge variety of system architectures to choose from, as we eventually develop interplanetary colonization. Such approach might even be more economically viable, if capitalism or a variation thereof is still in play when this happens.

Given that the amusement park would have to be a planet, it is only natural to have the parking lot as another separate body orbiting the star. A planet with its corresponding moon could be used for this purpose \footnote{Not to be confused with the future maximum security prison on the moon (Best et al. in prep.).}, unfortunately we know very little about exomoon populations. Also, the surface areas of both the amusement park and parking lot need to be approximately the same size (see Section 2.1); and, assuming roughly the same density for the planets, the system would consist of two planets orbiting each other (considering both structures would occupy the entire planetary surface, as to avoid being wasteful).

We also discard the possibility of building a separate artificial satellite, since this would be incredibly expensive, and because all you need to build a parking lot is a smooth surface. As such, a mega structure would be unnecessary allocation of funds. This leaves us with the option of having two different planets acting as an amusement park and a parking lot, separately.

The ideal configuration for these 2 planets would be the one where both planets are close enough in semi-major axis for quick transportation of tourists, but not too close for them to perturb each other and collide. Ideally, the planets should get to their closest approach once every year, assuming people still use the movement of Earth around the Sun to plan their vacations.

Another concern is that, unless humans have evolved beyond biological needs, we would need the planet to fall inside the habitable zone of its host star. This is the zone where water can exist in liquid form and it changes depending on the type of star we are considering. Furthermore, to make things more pleasant for future visitors, we would like to use the definition of the \textit{Really Habitable Zone} \citep{RHZ}, which is the zone where acceptable gin and tonics can exist (for parents, of course).

To further cut down on construction costs, we would like to have this system occur naturally, and so we compare it with actual observations of known exoplanetary systems (in Section 3.1). We have discovered so many of them, we need to put them to good use.

\section{Procedure} \label{procedure}

We ran simulations using the open-source N-body code Rebound \citep{rebound} with the following considerations, as to make the visitor feel at home.

\subsection{Parking lot surface area}

\begin{figure}[h!]
    \centering
    \includegraphics[width=8cm]{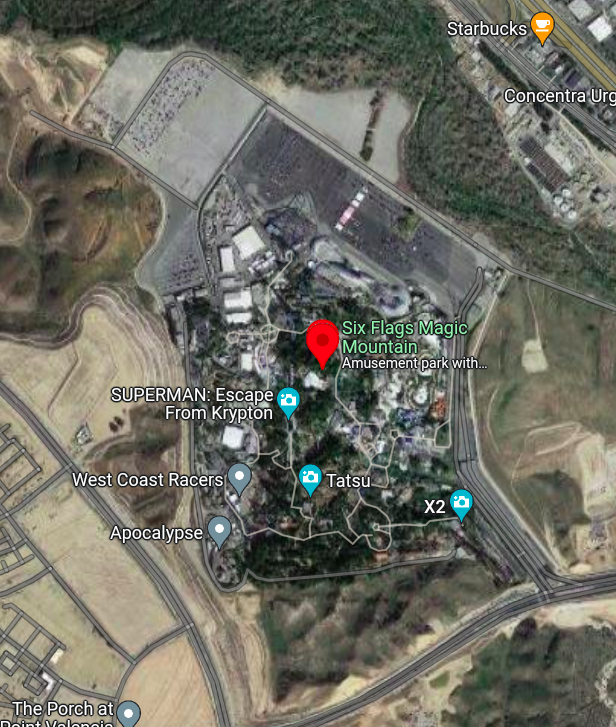}
    \caption{Satellite view of Six Flags Magic Mountain from Google Maps. The parking lot area can be seen in the top part of the picture and it is of a comparable size to the amusement park itself.}
    \label{fig:fig_sat_view}
\end{figure}

Because each family arrives in their own vehicle, the surface of the parking lots is often very similar to that of the buildings themselves\footnote{The actual calculation can be done as a back of the envelope estimation and it is left to the reader as an exercise.} (shopping centers, amusement parks, etc), although this fact is sometimes not noticeable because of underground parking lots. We include a satellite view of Six Flags Magic Mountain as an example in Fig. \ref{fig:fig_sat_view}. Note the presence of a Starbucks on the upper right corner of the satellite picture, confirming the theory \footnote{\href{http://www.markstivers.com/wordpress/comics/2005-01-01\%20Physics-of-c.gif}{http://www.markstivers.com/wordpress/comics/2005-01-01\%20Physics-of-c.gif}} that there is always a Starbucks \footnote{Not sponsored in any way... yet.} near you.

\subsection{Period of closest approach}

The perfect time for the transportation of passengers would be when the planets have their closest approach, as to avoid long travels with children. We can get this synodic period by estimating how much time it takes for the two planets to be 360 degrees apart from each other.

\begin{equation*}
    2 \pi \left( \frac{1}{P_1} - \frac{1}{P_2} \right)t = 2 \pi
\end{equation*}

\begin{equation}
    t = \frac{P_1 P_2}{P_2 - P_1}
\end{equation}

where $P_1$ and $P_2$ correspond to the period of the inner and outer planets, respectively.

\subsection{Really Habitable Zone (RHZ)}

In the same spirit as defining the Habitable Zone, a region of the planetary system where liquid water can exist, \cite{RHZ} did a very exhaustive study for the Really Habitable Zone, using the requirement of perfect gin and tonics. This translates to a requirement of effective stellar flux at the semimajor axis of the planet, and so the region depends on the type of star considered. We adopted constants from Table 1 in \cite{RHZ} and estimated this region for a star of mass 0.5$M_{\odot}$. This stellar mass gives the planets in this region an ideal synodic period of nearly 1 year, perfect for a summer vacation. We did not check if the planet would be tidally locked, but the presence of a permanent day side opens up the possibility of low cost tanning booths.

\subsection{Radius of the planet}

Given that we need a rocky surface for the planets to serve as construction grounds (specially for the parking lot), and also because thick atmospheres make takeoffs very difficult, we restrict ourselves to planets below a certain threshold in radius to ensure they are rocky planets. For a stellar mass of 0.5$M_{\odot}$, according to \cite{radiusgap}, this is close to 1.6 $R_{\oplus}$ or 4$M_{\oplus}$. Future companies might want to choose smaller planets, to consume less fuel during takeoff, but then their amusement park would have to be smaller too. There must be an ideal radius, but this is beyond the scope of the current paper and might be addressed on future works.

\subsection{Stability of the planetary system}

Thinking about the huge investment the construction on both planets would require, we would like for the planets not to collide and kill all the visitors, which would put us at huge risk of financial repercussions and lawsuits.

An analytic criterion for the critical spacing between two planets beyond which the system would be unstable is given in \cite{1993Icar..106..247G}. Defining
      \begin{equation}
          \Delta = \frac{a_2 - a_1}{R_H}
      \end{equation}
where  $a_1$ and $a_2$ are the semi-major axes of the two planets. \cite{1993Icar..106..247G} found a critical spacing of $\Delta_c = 2\sqrt{3} \approx 3.46$, where the mutual Hill radius ($R_H$) is given by:
     \begin{equation}
        R_H = \left(\frac{m_i + m_{i+1}}{3 M_s}\right)^{1/3}\left(\frac{a_i + a_{i+1}}{2}\right) \label{hill}
      \end{equation}
   
Next, since the stability of a two-planet system is determined by the well-studied three-body problem and the Gladman criterion seen above, we find it boring to just analyze a two-planet system. Considering that on average, according to \cite{ZhuPetrovich}, each Kepler-like system has 3 planets (N$_p$ =3), we ran some simulations (which justify the existence of this paper) to explore the stability of three-planet systems that fall under all of the conditions above.
    
For the Rebound simulations, the locations of the planets were also randomly set, making sure that at least 2 planets fall under the RHZ). Then, a third planet was added as a perturber on a wider range. As this perturber would not be a part of the amusement park, it is less constrained in mass and could even be a Neptune or a Jupiter. Actually, this planet could be used as a garbage disposal system if it had a sufficiently thick atmosphere to disintegrate all the trash from the amusement park's visitors.

\section{Results} \label{results}

\begin{figure}[h!]
    \centering
    \includegraphics[width=8cm]{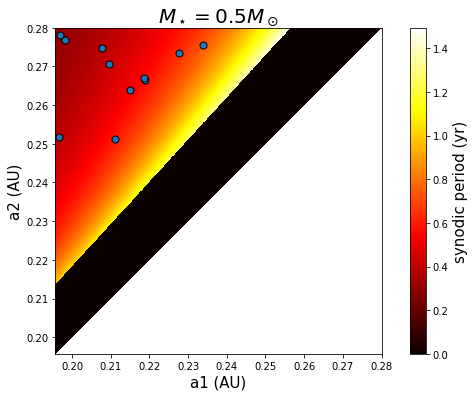}
    \caption{Systems which survived for 10 Myrs. Colors indicate the time between close approaches in years. The black band is prohibited by the Gladman criterion when there are only two planets. You do not want to be on the black band.}
    \label{fig:fig_sat}
\end{figure}

\begin{figure}[h!]
    \centering
    \includegraphics[width=7.5cm]{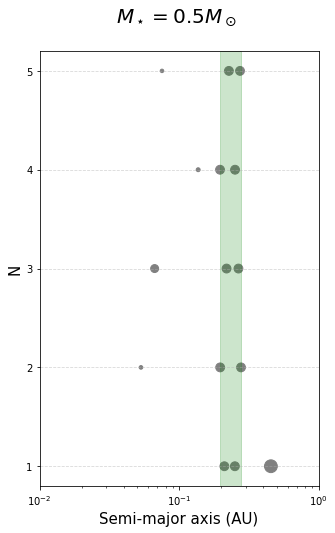}
    \caption{Some example systems from our simulations which survived at least for 10 Myrs. One of this could be your next vacation destination!}
    \label{fig:fig_sys}
\end{figure}

In Fig. \ref{fig:fig_sat} we show the planetary systems that survived at least for 10 Myr, which is plenty of time to recover our initial investment. We also indicate how long it will take for both planets in the RHZ to align for fast transportation, as to better schedule your vacations. No planetary system fell into the black band prohibited by the Gladman criterion, which is good for visitors who can now choose from our highly curated assortment of surviving systems showed in Fig. \ref{fig:fig_sys}.

\subsection{Comparing with nature}

\begin{figure}[h!]
    \centering
    \includegraphics[width=8cm]{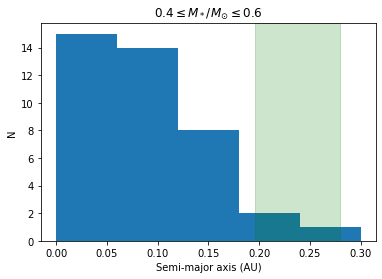}
    \caption{Basic python histogram of observed planets found near the stellar mass we considered. We also marked the RHZ so you know there are only 3 planets there.}
    \label{fig:fig_hist}
\end{figure}

Given that the RHZ is very narrow for the considered stellar mass, we could not find an illustrative exoplanetary system to match every requirement. Instead, we note that rocky planets for this stellar masses do occur (as seen in Fig. \ref{fig:fig_hist}), and hope for the next 5,000 planets.

\newpage

\section{Conclusions} \label{conclusions}

We have discussed the possibility of finding an ideal amusement park/parking lot planetary system (with room for an included garbage disposal planet) from a planetary architectural point of view. Our conditions for the system include: being at a perfect temperature for drinks, ideal transportation time between the planets, and long dynamical stability so you can come back as many times as you want. Further details of the planet's atmosphere, rotation period, and obliquity are left for future works. These might include the search for an ideal Coriolis effect for large scale water slides and possible snow parks at the poles.

We see that finding an ideal planet to include the parking lot is very challenging. Of course, this could all be avoided if a public transportation system were put in place to get visitors to the system. But we feel that, sadly, this is usually not the human way of doing things.

\acknowledgements

The authors would like to thank Cristobal Petrovich, Miguel Parra, Joaquin Silva and Eitan Dvorquez for insightful comments. If you want to do your own simulations and see if your 5 planet vacation resort could work, you can freely download Rebound at \href {http://github.com/hannorein/rebound}{http://github.com/hannorein/rebound}.

\bibliography{refs}

\end{document}